\documentclass[aps,prb,reprint,superscriptaddress,showpacs]{revtex4-1}
\pdfoutput=1
\usepackage{color}
\usepackage{graphicx}
\usepackage{amsmath}
\usepackage{amsfonts}
\usepackage{amssymb}
\usepackage{verbatim}
\usepackage{bbold}
\usepackage{hyperref}
\allowdisplaybreaks
\usepackage{multirow}
\usepackage{bbm}
\usepackage{braket}

\begin{document}

\author{Pieter~W.\ Claeys}
\email{PieterW.Claeys@UGent.be}
\affiliation{Institute for Theoretical Physics Amsterdam and Delta Institute for Theoretical Physics, University of Amsterdam, Science Park 904, 1098 XH Amsterdam, The Netherlands}
\affiliation{Department of Physics and Astronomy, Ghent University, Krijgslaan 281 S9, B-9000 Ghent, Belgium}
\affiliation{Center for Molecular Modeling, Ghent University, Technologiepark 903, 9052 Ghent, Belgium}

\author{Stijn~De\ Baerdemacker}
\affiliation{Department of Physics and Astronomy, Ghent University, Krijgslaan 281 S9, B-9000 Ghent, Belgium}

\author{Omar El Araby}
\affiliation{Institute for Theoretical Physics Amsterdam and Delta Institute for Theoretical Physics, University of Amsterdam, Science Park 904, 1098 XH Amsterdam, The Netherlands}

\author{Jean-S\'ebastien~Caux}
\affiliation{Institute for Theoretical Physics Amsterdam and Delta Institute for Theoretical Physics, University of Amsterdam, Science Park 904, 1098 XH Amsterdam, The Netherlands}

\title{Spin polarization through Floquet resonances in a driven central spin model }

\begin{abstract}
Adiabatically varying the driving frequency of a periodically-driven many-body quantum system can induce controlled transitions between resonant eigenstates of the time-averaged Hamiltonian, corresponding to adiabatic transitions in the Floquet spectrum and presenting a general tool in quantum many-body control. Using the central spin model as an application, we show how such controlled driving processes can lead to a polarization-based decoupling of the central spin from its decoherence-inducing environment at resonance. While it is generally impossible to obtain the exact Floquet Hamiltonian in driven interacting systems, we exploit the integrability of the central spin model to show how techniques from quantum quenches can be used to explicitly construct the Floquet Hamiltonian in a restricted many-body basis and model Floquet resonances.
\end{abstract}

\maketitle

\emph{Introduction.} -- Periodically driven systems have a rich history ranging from the simple kicked rotor to recent experimental progress on cold atoms in optical fields \cite{goldman_periodically_2014,bukov_universal_2015}. The dynamics in driven systems has remarkable features, such as the absence of a well-defined adiabatic limit \cite{hone_time-dependent_1997,weinberg_adiabatic_nodate} and the heating to infinite temperature which is expected to occur \cite{dalessio_long-time_2014,lazarides_equilibrium_2014, ponte_periodically_2015,else_pre-thermal_2017,moessner_equilibration_2017}. The same physical mechanism underlies these phenomena -- in the presence of periodic driving it is possible for states to interact resonantly. States whose energies are separated by an integer multiple of the driving frequency will interact strongly, leading to Floquet or many-body resonances \cite{eckardt_avoided-level-crossing_2008,hone_statistical_2009,bukov_heating_2016}. 

While this is generally seen as a disadvantage because of the experimental problems posed by heating, there is hope that in large but finite systems such many-body resonances can be well understood and even controlled. This opens up ways of engineering specific many-body resonant quantum states by adiabatically tuning the driving frequency to resonance. Such ``driven driving'' protocols, if smartly conceived, could even lead to states with properties beyond these associated with the (physical) driving Hamiltonians \cite{bendall_broadband_1995,hediger_adiabatic_1995,hwang_broadband_1997,takayoshi_laser-induced_2014,chen_floquet_2015,sato_laser-driven_2016}. This is illustrated on the central spin model, which is adiabatically driven such that the central spin becomes completely decoupled from its environment, incompatible with the physics of the instantaneous Hamiltonians. This model describes the (inhomogeneous) interaction of a central spin on which a magnetic field is applied with an environment of surrounding spins, being important in the study of quantum dots, solid-state nuclear magnetic resonance, and the nitrogen-vacancy defect in diamond, a promising qubit system \cite{balasubramanian_ultralong_2009}. A major experimental challenge remains the decoherence due to the presence of an environment, motivating numerous studies \cite{schliemann_electron_2003,yuzbashyan_solution_2005,bortz_exact_2007,bortz_spin_2007,erbe_different_2010,schliemann_spins_2010,bortz_dynamics_2010,barnes_nonperturbative_2012,stanek_dynamics_2013,faribault_integrability-based_2013,van_den_berg_competing_2014}. Surprisingly, Floquet resonances can here be used to construct pure spin states, seemingly at odds with the inevitable interaction with the environment and resulting decoherence effects.

All such resonances are encoded in the spectrum of the Floquet Hamiltonian, governing periodic dynamics. However, due to the exponential scaling of the Hilbert space and the inherently non-diagonal nature of time evolution operators, it is generally impossible to obtain this Hamiltonian in realistically-sized interacting systems. In the present paper we exploit that the system is driven by periodically switching between integrable Hamiltonians \cite{gaudin_bethe_2014,dukelsky_colloquium:_2004}, and we show how techniques from quantum quenches in integrability can be adapted to accurately model such transitions by constructing a (numerically) exact Floquet Hamiltonian in a restricted Hilbert space spanned by the resonant (Bethe ansatz) eigenstates of the integrable time-averaged Hamiltonian. This also presents a first step toward applying the toolbox from integrability to driven systems, where integrability is generally expected to lose its usefulness because of the general non-integrability of the Floquet Hamiltonian \cite{prosen_time_1998,claeys_breaking_2017,seetharam_absence_2017}. 

\emph{Floquet theory.} -- The key result in the study of periodically-driven systems is the Floquet theorem \cite{shirley_solution_1965}, recasting the unitary evolution operator as
\begin{equation}
{U}(t) = {P}(t) e^{-i {H}_F t},
\end{equation}
with ${P}(t)$ a periodic unitary operator with the same period $T$ as the driving, satisfying $P(T)=\mathbbm{1}$, and ${H}_F$ the Floquet Hamiltonian (with $\hbar=1$). Considering time evolution over one full cycle leads to the Floquet operator, from which $H_F$ can be extracted as
\begin{equation}
{U}_{F} \equiv {U}(T) = e^{-i {H}_F T}.
\end{equation}
Simultaneously diagonalizing these operators leads to
\begin{align}
{H}_F = \sum_n \epsilon_n \ket{\phi_n}\bra{\phi_n}, \ \ \ 
{U}_F = \sum_n e^{-i \theta_n} \ket{\phi_n}\bra{\phi_n},
\end{align}
with quasi-energies $\epsilon_n=\theta_n/T$. These provide the Floquet equivalent of quasi-momenta in Bloch theory, similarly only defined up to shifts $k \cdot 2 \pi/T, k \in \mathbb{N}$, and quasi-energies separated by shifts $k \cdot 2 \pi/T, k \in \mathbb{N}$ are said to be quasi-degenerate. Crucially, the Floquet Hamiltonian itself also depends on the driving period $T$. Consider a periodic quenching driving protocol
\begin{equation}
{H}(t) = 
  \begin{cases}
 {H}_1 &\text{for}\qquad 0 < t < \eta T, \\
  {H}_2    & \text{for}\qquad \eta T < t < T,
  \end{cases}
\end{equation}
with ${H}(t+T) = {H}(t)$ and $\eta \in [0,1]$, leading to 
\begin{equation}
{U}_F \equiv e^{-i {H}_F T} = e^{-i (1-\eta) T {H}_2} e^{-i \eta T {H}_1}.
\end{equation}
Obtaining the Floquet Hamiltonian from this expression is a non-trivial task, with exact results limited to systems where there is a clear commutator structure in all involved Hamiltonians \cite{gritsev_integrable_2017} (e.g. non-interacting systems \cite{russomanno_periodic_2012,lazarides_periodic_2014,russomanno_entanglement_2016}) or small systems for which exact diagonalization is feasible \cite{weinberg_quspin:_2017}. 


Still, the dependence of the Floquet Hamiltonian on the driving frequency has become well-understood in recent years \cite{dalessio_long-time_2014,bukov_heating_2016,weinberg_adiabatic_nodate}. At high driving frequencies, the Floquet Hamiltonian can be accurately approximated by an effective Hamiltonian, leading to strongly suppressed heating \cite{mori_rigorous_2016,kuwahara_floquetmagnus_2016,abanin_effective_2017,abanin_rigorous_2017}. This effective Hamiltonian can be obtained from the Magnus expansion, where the time-averaged Hamiltonian $H_{Avg} = \eta H_1 + (1- \eta) H_2$  presents a first-order approximation \cite{eckardt_high-frequency_2015,mikami_brillouin-wigner_2016,klarsfeld_baker-campbell-hausdorff_1989,blanes_magnus_2009}. Lowering the driving frequency $2 \pi/T$, many-body resonances are introduced where quasi-degenerate eigenstates of this effective Hamiltonian interact strongly and hybridize \cite{dalessio_long-time_2014,bukov_heating_2016,weinberg_adiabatic_nodate}. Further lowering the driving frequency, these many-body resonances multiply and lead to so-called `infinite-temperature states'. Here, it is crucial to consider finite systems with a bounded spectrum,  since the unbounded spectrum in infinitely large systems immediately leads to a proliferation of many-body resonances and `infinite temperature' Floquet eigenstates at all possible driving frequencies. However, by tuning the driving frequency in finite systems it remains possible to target specific resonances of non-trivial states, as will be illustrated.


\emph{The central spin model.} -- The model Hamiltonian is
\begin{equation}
{H} = B_z S_0^z + \sum_{j=1}^{L} A_j \vec{S}_0 \cdot \vec{S}_j,
\end{equation}
with $S_0^{\alpha}$ and $S_j^{\alpha}$ the spin operators of the central spin and the environment respectively. These are taken to be spin-$1/2$ particles, and the coupling constants are commonly chosen as $A_j = \exp\left[-(j-1)/L\right]$, corresponding to a quantum dot in a 2D Gaussian envelope \cite{coish_hyperfine_2004}. However, the integrability of the central spin model is versatile enough that our proposed method holds for arbitrary spins and parametrizations. For consistency with the literature on integrability, we set $\epsilon_j = -A_j^{-1}$ and $\epsilon_0=0$. The exact Bethe ansatz eigenstates
\begin{equation}
\ket{B_z;v_1 \dots v_N} = \prod_{a=1}^N \left(\sum_{j=0}^{L}\frac{S_j^+}{\epsilon_j-v_{a}}\right)\ket{\downarrow \dots \downarrow},
\end{equation}
depend on variables $\{v_1 \dots v_N\}$ satisfying Bethe equations
\begin{equation}
B_z^{-1} + \frac{1}{2}\sum_{j=0}^L\frac{1}{\epsilon_j-v_a} = \sum_{b \neq a}^N\frac{1}{v_b-v_a}, \qquad \forall a=1 \dots N,
\end{equation}
leading to energies
\begin{equation}
E(B_z;\{v_1 \dots v_N\}) = \frac{1}{2}\sum_{a=1}^Nv_a^{-1} - \frac{1}{4}\sum_{j=1}^L A_j-\frac{1}{2}B_z.
\end{equation}
Integrability now has two major advantages. First, these equations can be efficiently solved in a time scaling polynomially with system size. This should be contrasted with the conventional diagonalization of the Hamiltonian matrix in an exponentially large Hilbert space, allowing for exact results for large system sizes. Second, it allows for the systematic targeting of eigenstates through the Bethe equations. The key to our proposed approach is that overlaps between eigenstates of central spin Hamiltonians $\braket{B_{z,1};v_1 \dots v_N|B_{z,2};w_1 \dots w_N}$ with different magnetic fields $B_{z,1} \neq B_{z,2}$ can also be efficiently calculated numerically \footnote{See Supplemental Material for technical details, which includes Refs. [\onlinecite{slavnov_calculation_1989, korepin_quantum_1993,claeys_inner_2017,links_algebraic_2003,faribault_exact_2008,gorohovsky_exact_2011,babelon_bethe_2007,el_araby_bethe_2012,claeys_eigenvalue-based_2015,faribault_quantum_2009,Faribault2009,links_algebraic_2003}].}.

\begin{figure}
\includegraphics{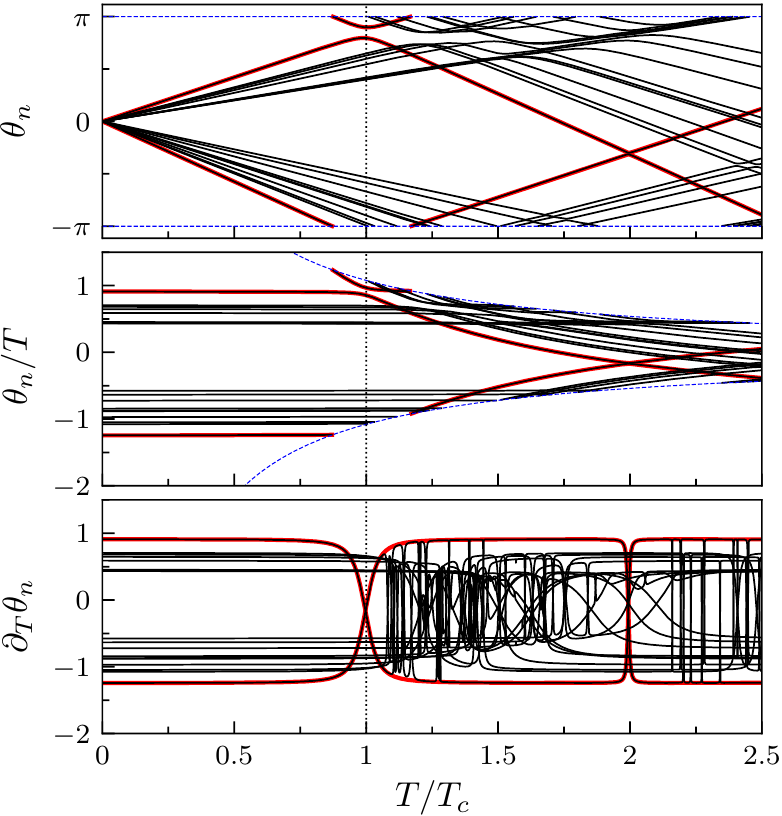}
\caption{Phase spectrum of the Floquet operator, quasi-energies and dynamical energies for a periodically driven central spin Hamiltonian at different driving periods $T$. The dotted (blue) lines mark the edges of the Brillouin zone $\pm \pi$ $(\theta_n)$ and $\pm \pi/T$ $(\theta_n/T)$, while the vertical dotted line denotes $T_c=2\pi/W$. The ground state and highest excited state are highlighted red using the approximative results from integrability (see further). \label{spec_centralspin}}
\end{figure}

Returning to Floquet dynamics, a protocol is considered where $B_z$ is periodically switched between values $B_{z,1}$ and $B_{z,2}$. To fix ideas, the eigenphases of the Floquet operator have been given in Figure \ref{spec_centralspin} for different driving periods $T$, with total spin projection 0, $\eta=0.5$ and $B_z$ switched between $1.2$ and $0.8$. These calculations have been performed using exact diagonalization on a small system with $L=5$ in order to avoid a visual clutter of eigenstates, but are representative for larger system sizes. Next to the spectrum of the Floquet operator, two energy measures of a Floquet state $\ket{\phi_n}$ are
\begin{equation}
\frac{\theta_n}{T}=\bra{\phi_n} {H}_F \ket{\phi_n}, \qquad \frac{\partial \theta_n}{\partial T} = \bra{\phi_n}{H}_{Avg} \ket{\phi_n},
\end{equation}
with $\theta_n/T$ the quasi-energies and $\partial_T \theta_n$ the dynamical contribution to the quasi-energies \cite{grifoni_driven_1998,claeys_breaking_2017}. This second quantity is convenient for the visualization of avoided crossings in the spectrum of the Floquet Hamiltonian.

At small driving periods, the spectrum of $H_F$ reduces to that of ${H}_{Avg}$ and both energies coincide. The onset of many-body resonances can be observed at $T_c = 2\pi/W$, with $W=E^{Avg}_{max}-E^{Avg}_{min}$ the bandwidth of ${H}_{Avg}$. At this critical frequency, the energy difference between the ground state and the highest excited state exactly matches the driving frequency. These states are then quasi-degenerate and interact resonantly, which can be clearly observed in the avoided crossing between their respective quasi-energies in $\braket{H_{F}}$ \footnote{After shifting one of the two states into the first Brillouin zone $[-\frac{\pi}{T},\frac{\pi}{T}]$} and the crossing between their respective energies in $\braket{H_{Avg}}$. Further increasing the driving period, more and more resonances are introduced. Remarkably, the off-resonant parts of the spectrum can often be accurately approximated using $H_{Avg}$ \cite{bukov_heating_2016,russomanno_floquet_2017}.

\emph{Resonant transitions.} -- Resonances have a major influence on the concept of adiabaticity, with distinct effects on the eigenstates of the Floquet Hamiltonian and the time-averaged Hamiltonian \cite{breuer_adiabatic_1989,young_adiabatic_1970,eckardt_superfluid-insulator_2005,russomanno_kibble-zurek_2016,russomanno_floquet_2017,russomanno_spin_2017}. Starting from an eigenstate of the Floquet Hamiltonian and adiabatically changing the driving frequency \footnote{Provided the micromotion operator does not change, see Ref. \onlinecite{weinberg_adiabatic_nodate}.}, the initial state will adiabatically follow the eigenstate of the Floquet Hamiltonian. Across resonance, this would result in a transition from e.g. the ground state to a highly excited state of the time-averaged Hamiltonian, since the eigenstates of the Floquet Hamiltonian adiabatically connect these states. 

\begin{figure}
\includegraphics{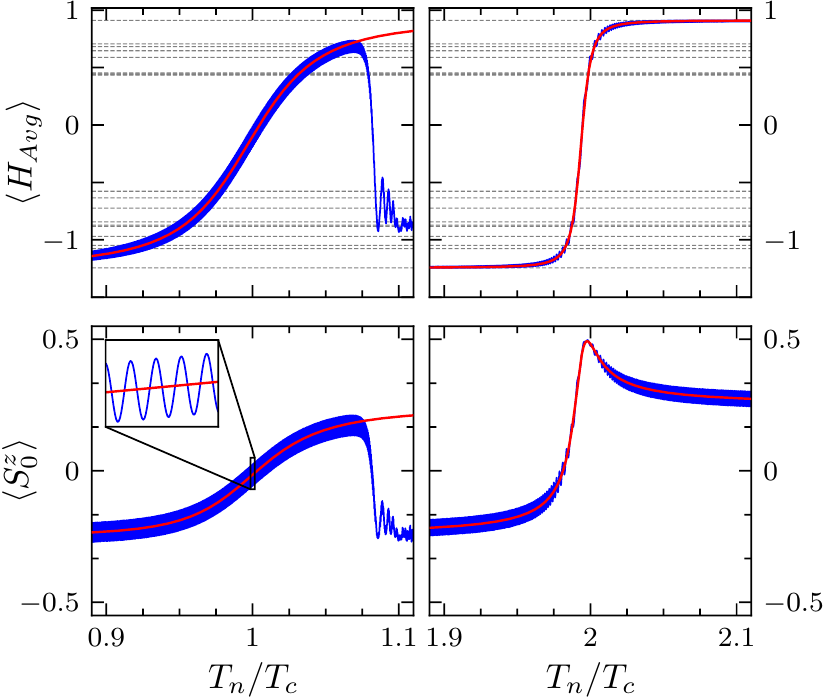}
\caption{Expectation values of ${H}_{Avg}$ and $S^z_0$ w.r.t. the adiabatic ground state of the Floquet Hamiltonian when the driving period is slowly increased from $0.8 \ T_c$ to $1.2 \ T_c$ (first column) or from $1.8 \ T_c$ to $2.2 \ T_c$ (second column) with $T_{i+1}-T_i = 10^{-4}$. Blue lines are exact results, while the red line is the approximation from integrability (see further). The dashed lines indicate the spectrum of $H_{Avg}$. \label{transition_L=6}}
\end{figure}

Focusing on the ground and highest excited state and adiabatically increasing the driving period across resonance, starting from the ground state of $H_{Avg}$ leads to
\begin{equation}
U(T_n) \dots U(T_2) U(T_1) \ket{\phi_0(\overline{B}_z)},
\end{equation}
with $T$ slowly increased from $T_1$ to $T_n$ and $\overline{B}_z = \eta B_{z,1}+(1-\eta) B_{z,2}$. We will refer to this state as the `adiabatic ground state', which is expected to adiabatically follow the corresponding eigenstate of the Floquet Hamiltonian through the frequency sweep, leading to transitions between resonant states. For the small system with $L=5$, such transitions are shown in Figure \ref{transition_L=6} for the first $(T \approx T_c)$ and second-order ($T \approx 2 T_c$) resonance. 

Slowly increasing the driving period, the system ends up in the highest excited state of ${H}_{Avg}$ in the second resonance, while it undergoes another resonance in the first transition before the highest excited state can be reached. Since the initial state is not an exact eigenstate of ${H}_F$, but only a (good) approximation, oscillations are introduced in all expectation values corresponding to contributions from excited eigenstates of $H_F$ to $\ket{\phi_0(\overline{B}_z)}$ (see inset in Figure \ref{transition_L=6}). These arise from higher-order contributions to the Magnus expansion and are as such controllable (e.g. by decreasing $|B_{z,2}-B_{z,1}|$). Still, it is clear that the ground state adiabatically follows the eigenstates of the Floquet Hamiltonian if the driving period is varied adiabatically. In order to have a clear transition between two states it is important that the resonance is isolated, where only a single state is quasi-degenerate with the ground state. For the ground and highest excited state, the first and second resonance are guaranteed to be isolated because of the two-band nature and the low density of states at the edge of the spectrum (see Figure \ref{spec_centralspin}). Note how $\langle S_0^z \rangle$, as shown in the lower panel of Figure \ref{transition_L=6}, vanishes at the first resonance and nears its maximal value of $1/2$ in the second resonance.

\emph{Modelling the resonant transition.} -- In general, such calculations require constructing the evolution operators for both driving Hamiltonians at each value of the driving period, and constructing and subsequently diagonalizing the Floquet operator. Each step involves the full Hilbert space, making such calculations unfeasible for realistic system sizes. However, knowledge acquired from quantum quenches (see e.g. [\onlinecite{jstatmech_outofequilibrium_2016}] and references therein) can be transferred to the present situation under the following key assumption. Namely, we assume that each many-body resonance can be modeled as a two-level system including only the corresponding quasi-degenerate eigenstates of $H_{Avg}$. This assumes that quasi-degenerate states do not interact strongly with off-resonant states or other quasi-degenerate states with a different quasi-energy, similar in spirit to degenerate perturbation theory. This approximation can be validated through the Floquet-Magnus expansion and is expected to hold if the deviations of the driving Hamiltonians are small w.r.t. the time-averaged Hamiltonian \cite{mikami_brillouin-wigner_2016,klarsfeld_baker-campbell-hausdorff_1989,blanes_magnus_2009,kuwahara_floquetmagnus_2016}. The Floquet operator can then be constructed in the $2$-dimensional basis $\{\ket{\phi_0(\overline{B}_z)},\ket{\phi_f(\overline{B}_z)}\}$ spanned by the relevant quasi-degenerate eigenstates of the time-averaged Hamiltonian 
\begin{equation}\label{UF}
{U}_F = 
\begin{bmatrix}
\braket{\phi_0(\overline{B}_z)|{U}_F|\phi_0(\overline{B}_z)} & \braket{\phi_0(\overline{B}_z)|{U}_F|\phi_f(\overline{B}_z)} \\
\braket{\phi_f(\overline{B}_z)|{U}_F|\phi_0(\overline{B}_z)} & \braket{\phi_f(\overline{B}_z)|{U}_F|\phi_f(\overline{B}_z)} 
\end{bmatrix}.
\end{equation}
Explicitly writing out the Floquet operator (\ref{UF}) and expanding in the eigenstates of the driving Hamiltonians, each matrix element is given by
\begin{align}\label{UF_mat}
\braket{\phi_i(\overline{B}_z)| {U}_F|\phi_j(\overline{B}_z)} =&  \sum_{m,n} e^{-i(1-\eta) E_m(B_{z,2}) T } e^{-i \eta E_n(B_{z,1}) T } \nonumber\\
& \times \braket{\phi_i(\overline{B}_z)|\phi_m(B_{z,2})} \nonumber\\
& \times \braket{\phi_m(B_{z,2})|\phi_n(B_{z,1})} \nonumber\\
& \times \braket{\phi_n(B_{z,1})|\phi_j(\overline{B}_z)}.
\end{align}
The calculation of each matrix element generally involves a double summation over the Hilbert space of energies and overlaps, which in turn involve summations over the Hilbert space. Integrability already provides numerically efficient expressions for the energies and the overlaps. As noticed in quantum quenches, another important feature of Bethe states is that they offer a basis in which only a very small minority of eigenstates carry substantial correlation weight, allowing summations over the full Hilbert space to be drastically truncated \cite{faribault_quantum_2009,Faribault2009}. Such a truncation scheme is presented in the Supplemental Material \cite{Note1}, and the induced error can be checked from sum rules. In practice, this allows for a numerically exact construction of the matrix elements (\ref{UF_mat}) for relatively large systems. 
\begin{figure}[t]
\includegraphics{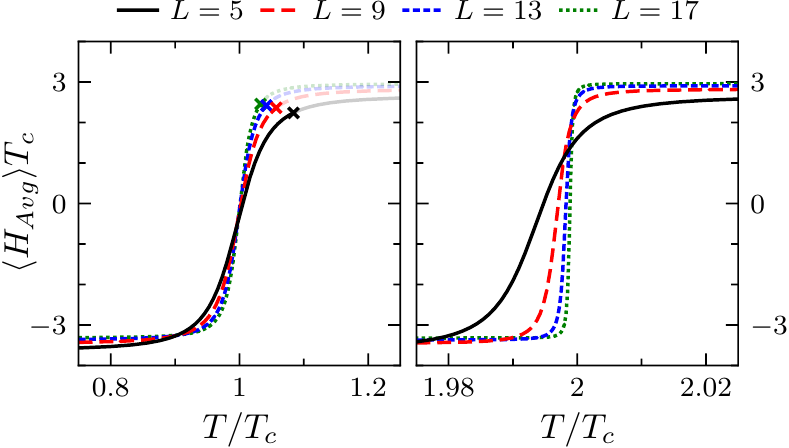}
\caption{Expectation value of the time-averaged Hamiltonian in the adiabatic ground state of the Floquet Hamiltonian with driving  $B_z=1 \pm 0.2$ and $\eta=1/2$ for different system sizes $L$. \label{compareL}}
\end{figure}

\begin{figure}[t]
\includegraphics{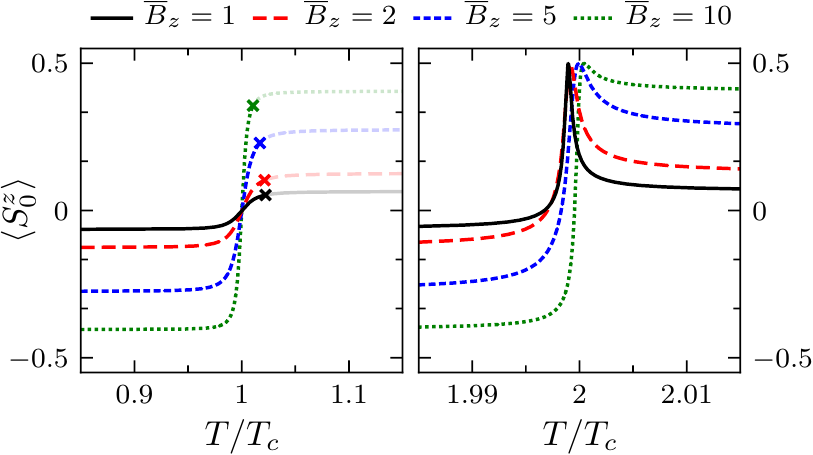}
\caption{Magnetization of the central spin in the adiabatic ground state of the Floquet Hamiltonian at different driving periods with $L=25$ and driving $B_z=\overline{B}_z \pm 0.2$. \label{compareB}}
\end{figure}

The resulting $2 \times 2$ operator can be easily diagonalized, and integrability allows for an efficient calculation of expectation values from its eigenstates \cite{slavnov_calculation_1989, korepin_quantum_1993,links_algebraic_2003,faribault_exact_2008,gorohovsky_exact_2011,claeys_inner_2017}. So the main approximation in this scheme is the restriction of the Hilbert space to a $2$-dimensional space, but within this space the Floquet operator is numerically exact. While we focus on the interaction between the ground and the highest excited state only, this makes it possible to systematically reconstruct part of the Floquet spectrum by including an increasing number of states in this basis. The accuracy can already be appreciated from Figures \ref{spec_centralspin} and \ref{transition_L=6}, where the avoided crossings near the resonances are well approximated but do not take into account the resonances involving other states. The results are extended in Figures \ref{compareL} and \ref{compareB} to different system sizes and average magnetic fields. The period beyond which the two-level approximation fails because another state needs to be included can also be estimated \cite{Note1} and is marked in both Figures. Note that this period lies outside the Figure for the second-order resonance.

\emph{Discussion.} --  While the expectation value of $H_{Avg}$ varies smoothly from the initial to the final value, the behaviour of the central spin is highly dependent on the order of the resonance. The magnetization $\braket{S_0^z}$ vanishes at the first resonance, while it nears the maximal value $1/2$ at the second resonance. Such a protocol could then be used to realize a state with magnetization exceeding that of both states, incompatible with any stationary central spin Hamiltonian, since a maximal value of $1/2$ implies a pure state decoupled from its environment.

A simple way to understand this behavior follows from the structure of the ground and the highest excited state, where the environment spins tend to align either anti-parallel or parallel to the central spin. These can be approximated by treating the environment as a single collective spin, and in this space the Hamiltonian simplifies to
\begin{equation}
{H} \approx \overline{B}_z S_0^z + A_b\vec{S}_b \cdot \vec{S}_0.
\end{equation}
Although not a necessary assumption \cite{Note1}, some intuition can be gained by taking $|\overline{B}_z| \ll |A_b S_b|$, where the relevant eigenstates can be approximated as
\begin{equation}\label{eigenstates}
\ket{\phi_{\pm}}\approx\frac{1}{\sqrt{2}} \left(\ket{\tfrac{1}{2},\tfrac{1}{2}}_0\ket{S_b,-\tfrac{1}{2}}_b \pm \ket{\tfrac{1}{2},-\tfrac{1}{2}}_0\ket{S_b,\tfrac{1}{2}}_b\right).
\end{equation}
At resonance, the Floquet states are approximately given by $\ket{\phi} = \frac{1}{\sqrt{2}} \left(\ket{\phi_{+}} \pm e^{i \theta} \ket{\phi_-}\right)$, where the relative phase $\theta$ is a priori unknown. However, the magnetization of the central spin depends on this relative phase as $\braket{\phi | S_0^z | \phi} = \frac{1}{2}\cos(\theta)$. The different magnetizations hence correspond to different relative phases acquired by these states. This relative phase can be deduced from second-order perturbation theory, expanding the matrix elements of the Floquet operator (\ref{UF_mat}) at resonance for small deviations from the average magnetic field $(B_{z}-\overline{B}_z)$ \cite{Note1}. Evolving either state over a full driving cycle will lead to a global phase and introduce off-diagonal corrections on the initial state, which are shown to either interfere constructively or destructively depending on the order of the resonance. This is reflected in the dependence of the off-diagonal elements on the order of the resonance $k$ through terms $e^{\pm i \eta k 2 \pi}$, and perturbation theory leads to relative phases $\pi/2$ and $3 \pi/2$ in the first resonance, while it leads to relative phases $0$ and $\pi$ in the second resonance. These explain the observed magnetization $\braket{S_0^z}=0$ or $\pm 1/2$ and the decoupling of the central spin. This behaviour extends towards higher-order resonances, where the polarization occurs at even-order resonances but vanishes at odd-order resonances. However, there is no guarantee that such resonances will be isolated and hence observable.

\emph{Conclusion.} -- In this work we investigated adiabatic transitions in the Floquet Hamiltonian when varying the driving frequency, leading to a transition between the ground and highest excited state away from resonance. Applying a periodically-varying magnetic field to a central spin model, it was shown how frequency sweeps and Floquet resonances can be used to prepare the system in a coherent superposition of the targeted states, either leading to a vanishing magnetization or a spin state exactly aligned with the magnetic field, depending on the order of the resonance. The latter effectively leads to a decoupling of the central spin from its environment, which can be used to purify the central spin. Integrability-based techniques were shown to be able to model this transition, which allows for an investigation of larger system sizes and presents a first step in applying techniques from integrability to interacting integrable systems subjected to periodic driving.


\emph{Acknowledgments.} -- We are grateful to M. Bukov, A. Russomanno, S.E. Tapias Arze and D. Van Neck for valuable discussions and/or comments. P.W.C. acknowledges support from a Ph.D. fellowship and a travel grant for a long stay abroad at the University of Amsterdam from the Research Foundation Flanders (FWO Vlaanderen). J.-S. C. acknowledges support from the Netherlands Organization for Scientific Research (NWO), and from the European Research Council under ERC Advanced grant 743032 DYNAMINT. This work is part of the Delta ITP consortium, a program of the Netherlands Organisation for Scientific Research (NWO) that is funded by the Dutch Ministry of Education, Culture and Science (OCW).

\bibliography{FloquetBib.bib}

\end{document}


\author{Pieter~W.\ Claeys}
\email{PieterW.Claeys@UGent.be}
\affiliation{Institute for Theoretical Physics Amsterdam and Delta Institute for Theoretical Physics, University of Amsterdam, Science Park 904, 1098 XH Amsterdam, The Netherlands}
\affiliation{Department of Physics and Astronomy, Ghent University, Krijgslaan 281 S9, B-9000 Ghent, Belgium}
\affiliation{Center for Molecular Modeling, Ghent University, Technologiepark 903, 9052 Ghent, Belgium}

\author{Stijn~De\ Baerdemacker}
\affiliation{Department of Physics and Astronomy, Ghent University, Krijgslaan 281 S9, B-9000 Ghent, Belgium}

\author{Omar El Araby}
\affiliation{Institute for Theoretical Physics Amsterdam and Delta Institute for Theoretical Physics, University of Amsterdam, Science Park 904, 1098 XH Amsterdam, The Netherlands}

\author{Jean-S\'ebastien~Caux}
\affiliation{Institute for Theoretical Physics Amsterdam and Delta Institute for Theoretical Physics, University of Amsterdam, Science Park 904, 1098 XH Amsterdam, The Netherlands}

\title{Spin polarization through Floquet resonances in a driven central spin model  -- Supplemental Material}

\pacs{}
\maketitle

\appendix
\section{Integrability of the central spin model}
\label{app:int}
The integrability of the central spin model is a crucial element in our analysis of the transitions in the Floquet quasi-energy spectrum. In this Appendix we provide the necessary technicalities in order to be self-contained. Following the notation from the main text, a set of operators can be defined as
\begin{equation}
Q_i = B_z S_i^z + \sum_{j \neq i}^L \frac{\vec{S}_i \cdot \vec{S}_j}{\epsilon_i-\epsilon_j},
\end{equation}
such that the central spin Hamiltonian corresponds to $Q_0$, and these operators mutually commute $[Q_i,Q_j]=0, \forall i,j$. These are the so-called conserved charges of the central spin model and can be simultaneously diagonalized by (unnormalized) Bethe ansatz states
\begin{equation}
\ket{B_z;v_1 \dots v_N} = \prod_{a=1}^N \left(\sum_{j=0}^{L}\frac{S_j^+}{\epsilon_j-v_{a}}\right)\ket{\downarrow \dots \downarrow},
\end{equation}
with rapidities $\{v_1 \dots v_N\}$ satisfying the Bethe equations
\begin{equation}
B_z^{-1} + \frac{1}{2}\sum_{j=0}^L\frac{1}{\epsilon_j-v_a} = \sum_{b \neq a}^N\frac{1}{v_b-v_a}, \qquad \forall a=1 \dots N.
\end{equation}
Following a famous result by Slavnov \cite{slavnov_calculation_1989, korepin_quantum_1993,claeys_inner_2017}, overlaps between Bethe states at different magnetizations can be calculated as
\begin{align}
&\braket{B_{z,1}; v_1 \dots v_N|B_{z,2};w_1 \dots w_N} \nonumber \\
&\qquad = \frac{\prod_{a} \prod_{b} (v_a-w_b)}{\prod_{b<a} (w_b-w_a) \prod_{a<b}(v_b-v_a) }  \det S_N,
\end{align}
with $S_N$ an $N \times N$ matrix defined as
\begin{align}
\left(S_N\right)_{ab} = \frac{1}{v_a-w_b}\bigg[&\sum_{j=0}^L \frac{1}{(v_a-\epsilon_j)(w_b-\epsilon_j)} \nonumber\\
&-2\sum_{c \neq a}^N \frac{1}{(v_a-v_c)(w_b-v_c)}  \bigg].
\end{align}
Taking the limit where the two sets of rapidities coincide leads to an expression for the normalization of a Bethe state as the determinant of a Gaudin matrix
\begin{equation}
\braket{B_{z};v_1 \dots v_N|B_{z};v_1 \dots v_N} =\det{G_N},
\end{equation}
with $G_N$ an $N \times N$ matrix defined as 
\begin{equation}
 \left(G_N\right)_{ab} =
  \begin{cases}
   \sum_{j=0}^L\frac{1}{(\epsilon_j-v_a)^2} -2 \sum_{c \neq a}^N \frac{1}{(v_c-v_a)^2}&\text{if}\ a=b \\
   \frac{2}{(v_a-v_b)^2}       & \text{if}\ a \neq b
  \end{cases}.
\end{equation}
This is sufficient for the evaluation of all terms in the Floquet operator. In order to calculate expectation values from the resulting wave function one more result needs to be used, namely that for the matrix elements of the spin projection $S_0^z$ (see e.g. Refs. [\onlinecite{links_algebraic_2003,faribault_exact_2008,gorohovsky_exact_2011}]). From the Hellmann-Feynman theorem, it follows that
\begin{align}
&\frac{\braket{B_{z}; v_1 \dots v_N|S_0^z|B_{z};v_1 \dots v_N}}{\braket{B_{z}; v_1 \dots v_N|B_{z};v_1 \dots v_N}} \nonumber \\
& \qquad \qquad \qquad \qquad = \frac{\partial }{\partial B_z}E(B_z;\{v_1 \dots v_N\}),
\end{align}
while the off-diagonal elements can be calculated as
\begin{align}
&\braket{B_{z}; v_1 \dots v_N|S_0^z|B_{z};w_1 \dots w_N} \nonumber \\
& \ \  = \frac{\prod_c (v_c-\epsilon_0)}{\prod_c(w_c-\epsilon_0)}\frac{\det\left({T_N+Q_N}\right)}{\prod_{b<a} (w_b-w_a) \prod_{a<b}(v_b-v_a) },
\end{align}
with $T_N$ following from $S_N$ as
\begin{equation}
\left(T_N\right)_{ab} = \left(S_N\right)_{ab} \prod_{c}(v_c-w_b),
\end{equation}
and $Q_N$ an $N \times N$ matrix defined as 
\begin{equation}
\left(Q_N\right)_{ab} = \frac{\prod_{c \neq b}(w_c-w_b)}{(\epsilon_0-v_a)^2} .
\end{equation}
Given a set of Bethe states, it is now possible to calculate all necessary overlaps and expecation values. Furthermore, the Bethe ansatz approach reduces the problem of obtaining eigenvalues and eigenstates to solving a set of coupled nonlinear equations, instead of diagonalizing the Hamiltonian in the full (exponentially large) Hilbert space. Many techniques have been proposed for solving these equations, where we have implemented the so-called eigenvalue-based approach \cite{babelon_bethe_2007,el_araby_bethe_2012,claeys_eigenvalue-based_2015}.
\vspace{\baselineskip}
\begin{figure}[ht]
\begin{center}
\includegraphics[width=0.9\columnwidth]{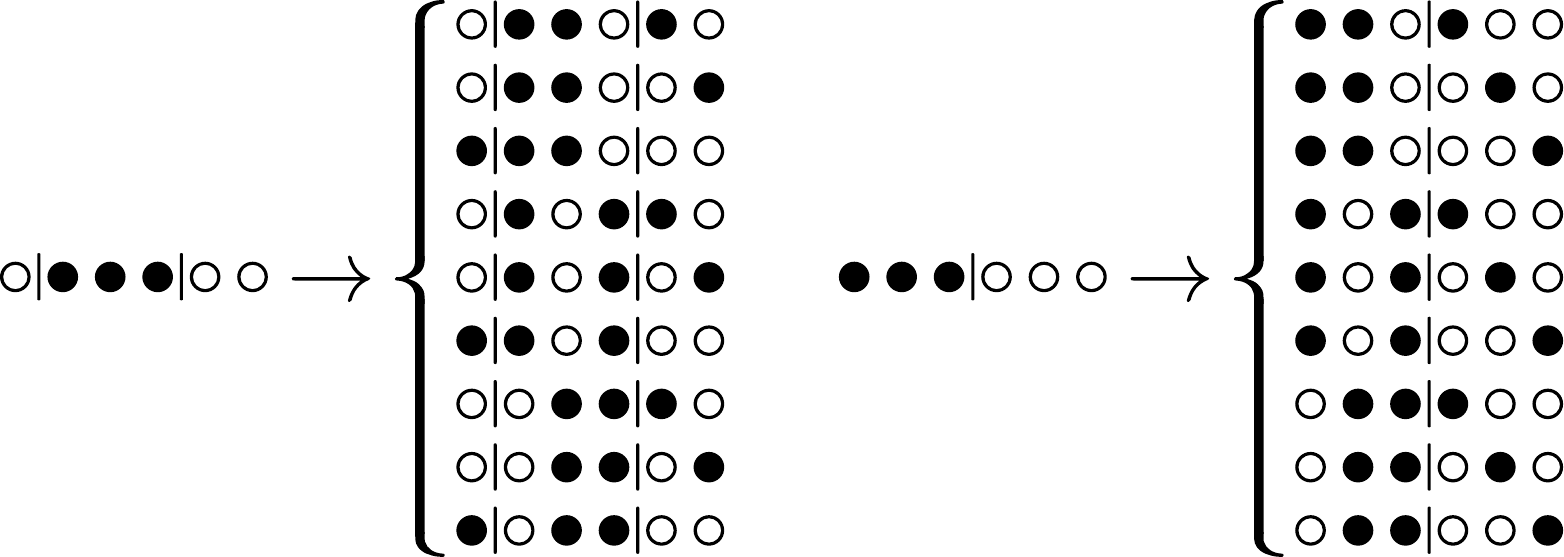}
\end{center}
\caption{Graphical representation of relevant states for $L=5$ and $N=3$ starting from the ground state and the highest excited state. \label{excitations}}
\end{figure}

The other necessary result for our approach was an efficient expansion of a Bethe state at given magnetic field $\bar{B}_z$ in a set of Bethe states at slightly different magnetic field. In Refs. \onlinecite{faribault_quantum_2009,Faribault2009}, it was shown how such an expansion can be efficiently implemented by targeting relevant eigenstates in a systematic way. This targeting can be represented in a graphical way by making the connection with the $B_z \to \infty$ limit. In this limit, the rapidities behave as $v_a = \epsilon_{i(a)} + \mathcal{O}(B_z^{-1})$, where $i(a)$ associates a spin index with each rapidity such that the eigenstates reduce to
\begin{equation}
\ket{B_z \to \infty;v_1 \dots v_N} \propto \prod_{a=1}^N S^+_{i(a)}\ket{\downarrow \dots \downarrow}.
\end{equation}
In this limit, all states reduce to simple product states defined in terms of a set occupied spin levels $\{i(1) \dots i(N)\}$. For the given parametrization, the ground state reduces to the state where the central spin is unoccupied and the states interacting most strongly with the central spin are occupied, which can be represented graphically (for e.g. $L=5$ and $N=3$) as
\begin{align}
\ket{\phi_0(B_z \to \infty)} &\propto S_1^+ S_2^+ S_3^+  \ket{\downarrow \dots \downarrow}\nonumber \\
&\qquad = \ket{\circ | \bullet \bullet \bullet | \circ \circ},  \\
\ket{\phi_f(B_z \to \infty)} &\propto S_0^+ S_1^+ S_2^+  \ket{\downarrow \dots \downarrow}\nonumber \\
&\qquad = \ket{\bullet \bullet \bullet | \circ \circ \circ}.
\end{align}
When expanding an eigenstate at fixed $B_z$ into eigenstates of a model at different $B_z$, it is often sufficient to restrict the expansion to states which can be related to the initial state by simple spin-flip excitations in the limit $B_z \to \infty$. This results in a set of $N(L+1-N)$ states for which the overlap needs to be calculated, which are again represented graphically in Figure \ref{excitations}.

When expanding an initial state $\ket{\phi_i(\bar{B}_z)}$ in such a restricted basis, a measure for the error is easily calculated as $1-\sum_{n}|\braket{\phi_i(\bar{B}_z)|\phi_n}|^2$, which reduces to zero for a complete basis. If this initial truncation would prove to be insufficient and the error exceeds a certain threshold (when e.g. there are large differences between the time-averaged Hamiltonian and the driving Hamiltonians), this summation can be extended in a systematic way by including higher-order spin-flip excitations.

\section{Perturbation expansion of the Floquet operator}
\label{app:PT}

\begin{figure*}[htb!]                      
 \begin{center}
 \includegraphics[scale=1]{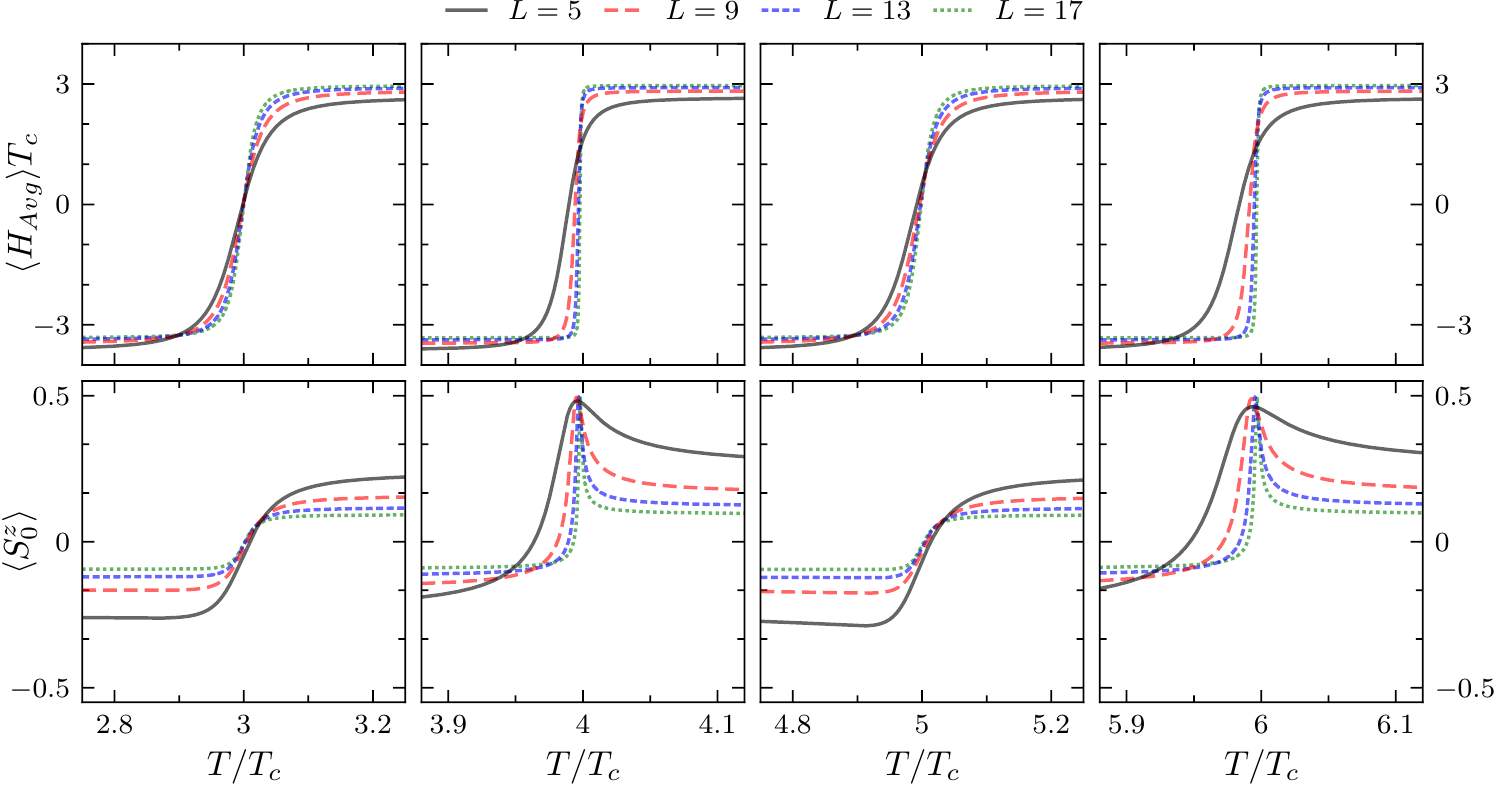}
 \caption{Expectation value of the time-averaged Hamiltonian $\braket{H_{Avg}}$ and magnetization of the central spin $\braket{S_0^z}$ in the adiabatic ground (and highest excited) state of the Floquet Hamiltonian at higher-order resonances with driving  $B_z=1 \pm 0.2$ and $\eta=1/2$ for different system sizes $L$. \label{fig:higherresonances}}
 \end{center}
\end{figure*}

In order to better understand the behaviour of the Floquet eigenstates near resonance, we perform a perturbative expansion of the Floquet operator when the model is being driven in such a way that there are only small deviations from the average magnetic field. Then within each matrix element all non-diagonal overlaps in the summation will be of order $B_{z,i} - \overline{B}_z \equiv \mathcal{O}(\Delta)$, allowing the summation to be severely restricted. For corrections up to $\mathcal{O}(\Delta^2)$, only the initial and final state are relevant as intermediate states in the summation. The diagonal elements can easily be found as
\begin{align}
\braket{\phi_0(\overline{B}_z)|{U}_F|\phi_0(\overline{B}_z)} &= e^{-i k E_0(\overline{B}_z)T_c} + \mathcal{O}(\Delta^2), \\
\braket{\phi_f(\overline{B}_z)|{U}_F|\phi_f(\overline{B}_z)} &= e^{-i k E_f(\overline{B}_z)T_c} + \mathcal{O}(\Delta^2),
\end{align}
which holds for arbitrary values of the driving period. In the first element, the summation has been restricted to $(m,n)=(0,0)$, while it has been restricted to $(m,n)=(f,f)$ in the second element. Here $0$ and $f$ label the resonant states, which we will take to be the ground (initial) state and highest excited (final) state. All other terms involve at least two off-diagonal overlaps and are as such of $\mathcal{O}(\Delta^2)$. Due to the expansion around $\overline{B}_z$, the first-order corrections on the phases can also be shown to vanish. Note that we discard a summation of $\mathcal{O}(\Delta^2)$ over an exponentially large Hilbert space, but we can assume that the phases do not add coherently and as such this summation can still be neglected. If this would not be the case then the resonant states have a strong interaction with off-resonant states and our 2-level approximation would similarly prove to be insufficient. 

The off-diagonal elements contain at least one off-diagonal overlap, restricting the summation to $(m,n) = (0,0)$, $(f,f)$ and $(0,f)$ or $(f,0)$, leading to
\begin{align}
&\braket{\phi_0(\overline{B}_z)|{U}_F|\phi_f(\overline{B}_z)} \nonumber \\
&= e^{-i(1-\eta) E_0(\overline{B}_z) T }e^{-i \eta E_0(\overline{B}_z) T } (B_{z,1}-\overline{B}_z)\braket{\partial_{B_z}\phi_0|\phi_f} \nonumber \\
&+ e^{-i(1-\eta) E_f(\overline{B}_z) T }e^{-i \eta E_f(\overline{B}_z) T } (B_{z,2}-\overline{B}_z) \braket{\phi_0|\partial_{B_z}\phi_f} \nonumber \\
& + e^{-i(1-\eta) E_0(\overline{B}_z) T }e^{-i \eta E_f(\overline{B}_z) T } \nonumber \\
& \ \ \times \left[(B_{z,2}-\overline{B}_z)\braket{\partial_{B_z}\phi_0|\phi_f} + (B_{z,1}-\overline{B}_z)\braket{\phi_0|\partial_{B_z}\phi_f}\right] \nonumber \\
& +\mathcal{O}(\Delta^2),
\end{align}
and similar for $\braket{\phi_0(\overline{B}_z)|{U}_F|\phi_f(\overline{B}_z)}$. Here the inner products have been expanded as e.g. $\ket{\phi_{0}({B}_{z,i})} =\ket{\phi_{0}(\overline{B}_{z})} + (B_{z,i}-\overline{B}_z)\ket{\partial_{B_z}\phi_{0}(\overline{B}_z)} + \mathcal{O}(\Delta^2)$, where the dependence on $\overline{B}_z$ has been made implicit in the final expressions. Evaluating these at $T= k \cdot T_c, k \in \mathbb{N}$ and explicitly setting $E_f = E_0 + 2\pi/T_c$, the matrix elements can (up to $\mathcal{O}(\Delta^2)$) be rewritten as
\begin{align*}
\braket{\phi_0(\overline{B}_z)|{U}_F|\phi_0(\overline{B}_z)} &= e^{-i k E_0(\overline{B}_z)T_c}, \\
\braket{\phi_f(\overline{B}_z)|{U}_F|\phi_f(\overline{B}_z)} &= e^{-i k E_F(\overline{B}_z)T_c}= e^{-i k E_0(\overline{B}_z)T_c}, \\
\braket{\phi_0(\overline{B}_z)|{U}_F|\phi_f(\overline{B}_z)} &= e^{-i k E_0(\overline{B}_z)T_c} (B_{z,1}-B_{z,2})   \nonumber \\
& \qquad \times \braket{\partial_{B_z} \phi_0|\phi_f} (1-e^{-i \eta k 2 \pi})   , \\
\braket{\phi_f(\overline{B}_z)|{U}_F|\phi_0(\overline{B}_z)} &= e^{-i k E_0(\overline{B}_z)T_c} (B_{z,1}-B_{z,2}) \nonumber \\
& \qquad  \times \braket{\partial_{B_z} \phi_f|\phi_0} (1-e^{+ i \eta k 2 \pi}).
\end{align*}
For the central spin model, it is known that all overlaps are purely real \cite{links_algebraic_2003} and hence $\braket{\partial_{B_z} \phi_0|\phi_f}=-\braket{\partial_{B_z} \phi_f|\phi_0}$. Taking $k=1$ and $\eta=1/2$ then results in 
\begin{align}
&U_F = \ e^{-i E_0(\overline{B}_z)T_c} \nonumber \\
&\ \times \left(\mathbbm{1} + 2 (B_{z,1}-B_{z,2}) \braket{\partial_{B_z} \phi_0|\phi_f}
\begin{bmatrix}
0 & 1\\
-1 & 0
\end{bmatrix}
\right) 
+ \mathcal{O}(\Delta^2).
\end{align}
Diagonalizing the $2 \times 2$ perturbation matrix leads to eigenstates 
\begin{equation}
\ket{\phi_0(\overline{B}_z)} \pm i \ket{\phi_f(\overline{B}_z)},
\end{equation}
corresponding to relative phases of $\pi/2$ or $3\pi/2$, as mentioned in the main text.

For the second resonance $(1-e^{\pm i \eta k 2 \pi})=0$ and the first-order correction vanishes, so higher-order terms need to be included. It is no longer possible to explicitly obtain the corrections without performing the summation over the Hilbert space, but symmetry arguments can be used to predict the relative phase. If the first-order correction vanishes, the Floquet operator can be written as
\begin{equation}
U_F = e^{-i E_0(\overline{B}_z) T} \left(\mathbbm{1} + (B_{z,1}-B_{z,2})^2 V(\overline{B}_z)  + \mathcal{O}(\Delta^3)\right),
\end{equation}
with $V(\overline{B}_z)$ a $2 \times 2$ matrix containing the second-order corrections on the matrix elements. Demanding $U_F$ to be unitary then results in the constraint that $V(\overline{B}_z)$ is anti-hermitian $V^{\dagger}(\overline{B}_z)=-V(\overline{B}_z)$. Hence, taking the transpose of $U_F$ with $\eta=1/2$ and exchanging $B_{z,1}$ and $B_{z,2}$ leaves $U_F$ invariant, since
\begin{align}
\left(U_F(B_{z,1},B_{z,2})\right)^T &=  \left(e^{-i T {H}(B_z,2)/2 } e^{-i \eta T {H}(B_{z,1})/2}\right)^T \nonumber \\
& = e^{-i T {H}(B_z,1)^T/2 } e^{-i \eta T {H}(B_{z,2})^T/2} \nonumber \\
& = e^{-i T {H}(B_z,1)/2 } e^{-i \eta T {H}(B_{z,2})/2} \nonumber \\
& = U_F(B_{z,2},B_{z,1}).
\end{align}
This restricts $V(\overline{B}_z)$ to be symmetric, since exchanging $B_{z,1}$ and $B_{z,2}$ leaves the second-order contribution $(B_{z,1}-B_{z,2})^2 V(\overline{B}_z)$ invariant. Combining these restrictions allows this matrix to be rewritten as $V(\overline{B}_z) =i W(\overline{B}_z) $, with $W(\overline{B}_z)$ a purely real and symmetric matrix. Diagonalizing this perturbation matrix, the eigenstates will be purely real and result in relative phases of $0$ or $\pi$.

Perturbation theory thus explains the relative phases in both resonances, where it is important to note that the presented arguments can be extended to systems where the condition $|B_z| \ll |A_b S_b|$ does not hold. The orthogonal eigenstates of $H_{Avg}$ can be approximately constructed as
\begin{align}
\ket{\phi_0}=&\cos (\varphi) \ket{\tfrac{1}{2},\tfrac{1}{2}}_0\ket{S_b,-\tfrac{1}{2}}_b + \sin (\varphi)\ket{\tfrac{1}{2},-\tfrac{1}{2}}_0\ket{S_b,\tfrac{1}{2}}_b, \nonumber \\
\ket{\phi_f}=&\sin (\varphi)\ket{\tfrac{1}{2},\tfrac{1}{2}}_0\ket{S_b,-\tfrac{1}{2}}_b - \cos (\varphi) \ket{\tfrac{1}{2},-\tfrac{1}{2}}_0\ket{S_b,\tfrac{1}{2}}_b. \nonumber \\
\end{align}
In the first-order resonance, it can easily be checked that a coherent superposition of these states with a relative phase of $\pi/2$ and $3\pi/2$ ($\ket{\phi_0}\pm i \ket{\phi_f}$) always has a vanishing magnetization. In the second-order resonance, the relative phases $0$ or $\pi$ result in Floquet eigenstates with real coefficients interpolating between these two states. This then necessitates a change of sign in one of the two components (either when going from $+\cos (\varphi)$ to $+\sin (\varphi)$ or from $+\sin (\varphi)$ to $-\cos (\varphi)$), resulting in an intermediate state where only a pure state remains and the central spin is decoupled. While this might not always be exactly at the point of resonance where $T=2T_c$, this will generally occur close to resonance because of the second-order nature of the perturbation. 

Perturbation theory then generally predicts odd-order resonances where the magnetization is approximately zero and even-order resonances near which the magnetization is either maximal or minimal, as supported by numerical results from the integrability-based approximation. This corresponds to the behaviour at first- and second-order resonances as discussed in the main text, and is shown in Figure \ref{fig:higherresonances} for the first four higher-order resonances. It should be kept in mind that there is no guarantee that these higher-order resonances will be isolated, since other states might be quasi-degenerate with the resonant states and the two-level approximation would fail. However, if the resonance only involves the two targeted states, the described dependence of the magnetization on the order of the resonance is expected to hold.

\section{Range of validity of the approximation}
As shown in the main text, the two-level approximation fails when other resonances occur. Starting from the ground state of $H_{Avg}$ and increasing the driving period across the first-order resonance with the highest excited state, the range of periods for which no other resonances are expected to occur can be estimated. Specifically, the next relevant resonance occurs when the highest excited state is quasi-degenerate with the first excited state. This state has energy $E^{Avg}_{min}+E^{Avg}_{gap}$, with $E^{Avg}_{gap}$ the ground-state energy gap in the time-averaged Hamiltonian, and a resonance is expected at $T = 2 \pi /(E^{Avg}_{max}-E_{min}^{Avg}-E^{Avg}_{gap})$. This is the relevant period discussed in the main text, leading to
\begin{equation}
\frac{T-T_c}{T_c} = \frac{E^{Avg}_{gap}}{E^{Avg}_{max}-E^{Avg}_{min}-E^{Avg}_{gap}} \approx \frac{E^{Avg}_{gap}}{W}.
\end{equation}
The relevant state can again be efficiently targeted from the $B_z \to \infty$ limit and is related to the ground state through a simple spin-flip excitation (e.g. for the previously mentioned example $\ket{\circ | \bullet \bullet \circ | \circ \bullet}$). This allows the energy gap to be easily calculated, returning the results mentioned in the main text. For the chosen parametrization the bandwidth $W$ scales linearly with system size $L$, whereas the energy gap $E^{Avg}_{gap}$ quickly converges to a constant (non-zero) value with increasing system size $L$. The relevant range of periods for which no additional resonances are expected to occur then scales inversely with system size since $E^{Avg}_{gap}/{W} \propto L^{-1}$. For the second-order resonance, this region can similarly be shown to be approximately $2 E^{Avg}_{gap}/W$. 

Two effects hence combine to lead to a second-order resonance which is more isolated compared to the first-order one -- it is a second-order perturbative effect and hence more narrow, and the subsequent region where no additional resonances are expected to occur is larger.

\bibliography{FloquetBib.bib}